# Improving the Closed-Loop Tracking Performance Using the First-Order Hold Sensing Technique with Experiments

Chifu Yang[1], Shuang Gao[2], Zhu Xue[3]

[1]Harbin Institute of Technology, Harbin, 150006, China
[2] McGill University, Montreal, Quebec, H3A 0E9, Canada
[3]National University of Singapore, 117576, Singapore

**Abstract**
This paper proposes a new perspective in the enhancement of the closed-loop tracking performance by using the first-order hold (FOH) sensing technique. Firstly, the literature review and fundamentals of the FOH are outlined. Secondly, the performance of the most commonly used zero-order hold (ZOH) and that of the FOH are compared. Lastly, the detailed implementation of the FOH on a pendulum tracking setup is presented to verify the superiority of the FOH over the ZOH in terms of the steady state tracking error. The results of the simulation and the experiment are in agreement.

## 1. Introduction

The first-order-hold (FOH) method is a mathematical model to reconstruct the sampled signals that could be done by a conventional digital-to-analog converter (DAC) and an analog circuit which is called an integrator. The FOH signal is reconfigured as a piecewise linear approximation of the original sampled signal. In [1], the FOH signal is used for real-time substructure testing, which is a novel method of testing structures under dynamic loading. An extrapolation of a first-order-hold discretization is used which increases the accuracy of the numerical model over more direct explicit methods. The improvements are demonstrated using a series of substructure tests on a simple portal frame. Some deep mathematical relationship between ZOH and FOH is revealed in [2-5]. It is shown that the zeros of sampled-data systems resulting from rapid sampling of continuous-time systems preceded by a ZOH are the roots of the Euler-Frobenius polynomials. The simplicity, negative realness, and interlacing properties of the sampling zeros of ZOH and FOH sampled systems are proven for the first time in literature. The paper [6] deals with the quality requirements of the synthesized sine waves reconstructed through a ZOH and FOH for testing purposes, especially when a switching demodulator is used. Results show that a FOH implies a decrease of total harmonic amplitude distortion, but the measured spurious harmonics are kept lower or equal when using a ZOH in the 15 closest components. It is concluded in the paper that in testing applications a ZOH yields better results, thus the benefits of using a FOH need further investigation. The effects of various sensing techniques on the performance of the motion canceling bilateral control (MCBC) are studied [7-9]. MCBC is a method to synchronize motion of a teleoperation robot and a target, while an operator can obtain tactile sensation of the remote target [10, 11]. Results show that a FOH yields better performance compared with a ZOH, but it has a peak gain near the Nyquist frequency. Therefore, in order to make full use of FOH, additional techniques are needed in order to eliminate the adverse effects of the FOH. The mathematical structure of new discretization schemes are proposed and characterized as useful methods of establishing concrete connections between numerical and system theoretical properties [12-14]. The paper [15] deals with the necessary and sufficient condition for the reachability of the sampled-data system obtained by the discretization of a linear time-invariant continuous-time system with a FOH. The equivalence of the reachability and controllability of the system is shown and similar results are given for observability and re-constructability [16-19].

## 2. First-order Hold

The motivation in this paper is on the better estimation of the analog signal based on the digital signal $\theta_{A/D}$ read from the analog-to-digital converter (ADC). As a result, the closed-loop tracking performance could be improved. For this, the first-order-hold (FOH) is a better method to approximate the continuous analog signal than the zero-order hold (ZOH) [20]. As mentioned previously, the FOH signal is a reconstructed piecewise linear approximation to the original sampled signal (see Figure 1).

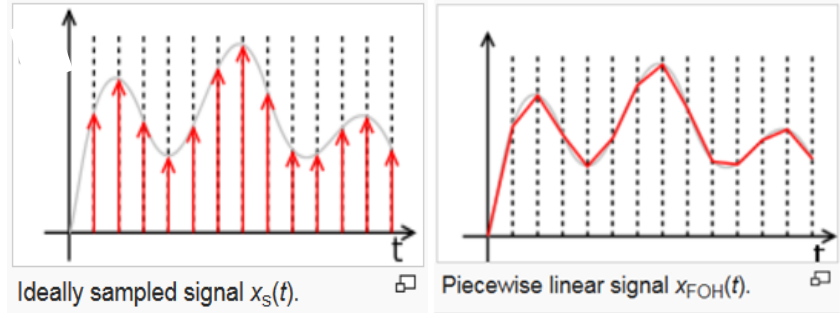

**Figure 1 [21].** Ideally sampled signal and corresponding piecewise linear FOH signal.

The ideally sampled signal could be represented as,
$$x_s(t) = T \sum_{n=-\infty}^{\infty} x(nT)\delta(t - nT) \qquad (1)$$
where $x(t)$ is the original signal, $x_s(t)$ is the ideal signal, $\delta(t)$ is the Dirac impulse function. Since a sequence of Dirac impulses, representing the discrete samples, is low-pass filtered, the mathematical model for FOH is necessary. The impracticality of outputting a sequence of Dirac impulses foster the development of devices that use a conventional DAC and some linear analog circuitry, to reconstruct the piecewise linear output for the FOH signals. The commonly used analytical piecewise linear approximation is written as,
$$x_{FOH}(t) = \sum_{n=-\infty}^{\infty} x(nT) tri\left(\frac{t-nT}{T}\right) \qquad (2)$$
where $tri$ is the triangular function defined as,
$$tri(t) = \max(1 - |t|, 0) \qquad (3)$$
However, the system represented in (2) is not achievable in realty. In fact, the typical FOH model used in practice is the delayed first-order hold, which is identical to the FOH except for the fact that its output is delayed by one sample period, resulting in a delayed piecewise linear output signal (see Figure 2).

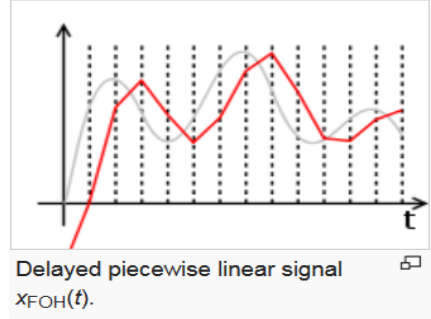

Delayed piecewise linear signal
$x_{FOH}(t)$.

**Figure 2 [21].** Delayed piecewise linear FOH signal

The delayed first-order hold, also known as causal first-order hold, as shown in Figure 2 can be represented as,

$$x_{FOH}(t) = \sum_{n=-\infty}^{\infty} x(nT)tri(\frac{t-T-nT}{T}) \quad (4)$$

The delayed output renders the system a causal system [22-25]. The corresponding delayed piecewise linear reconstruction is physically realizable with the assistance of a digital filter[26-28].

**2.1 First-order Hold VS Zero-order Hold**

The zero-order hold (ZOH) is a mathematical representation of the practical signal reconstruction done be a conventional digital-to-analog converted (DAC). It works in a way that the signal is held at each sample value for each and every sample interval while converting a discrete-time signal to a continuous-time signal. The most commonly used sensing feature in practice is the ZOH due to its ease of implementation [29-32]. The mathematical model of the ZOH is written as,

$$x_{ZOH}(t) = \sum_{n=-\infty}^{\infty} x[n]rect(\frac{t-nT}{T} - \frac{1}{2}) \quad (5)$$

where x[n] is the discrete samples, $rect()$ is the rectangular function as follows,

$$rect(t) = \begin{cases} 0, & if\ |t| > 1/2 \\ \frac{1}{2}, & if\ |t| = 1/2 \\ 1, & if\ |t| < 1/2 \end{cases} \quad (6)$$

Next, the properties of the FOH and the ZOH are compared as shown in Figure 3,

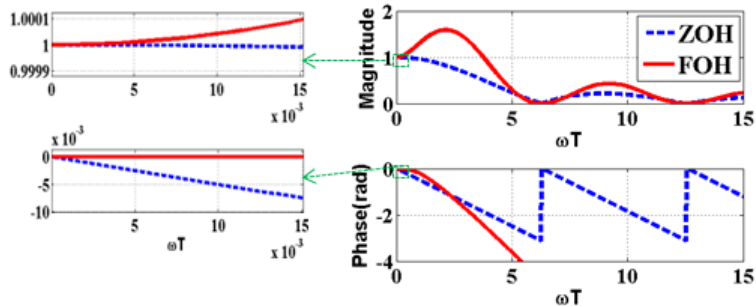

**Figure 3.** Magnitude and phase of ZOH and FOH filters

From Figure 3, for low frequency (below $\pi/2T$) signals, although the FOH has larger amplitude distortion than the ZOH does, the FOH has significantly less phase lag than the

ZOH does [20]. This property is utilized in this paper to reduce the level of steady state tracking error based on the fact that a more precise sensing signal is being utilized for the feedback. The details of the implementation are illustrated in the following section.

## 3. Implementation of the FOH on an Experimental Setup

The enhanced tracking performance while using the FOH instead of the ZOH is demonstrated over a single axis manipulator test platform (an under-actuated pendulum) as shown in Figure 4.

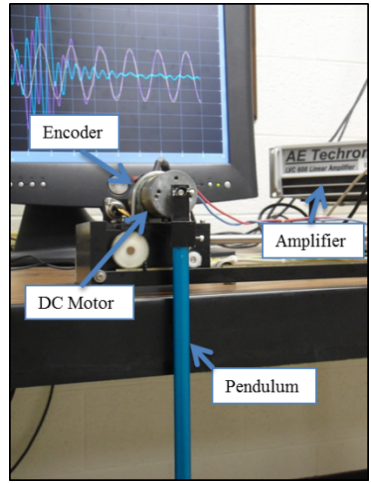

**Figure 4.** Experimental setup

The trajectory tracking performance of the pendulum is investigated. The desired trajectory is selected as a sinusoidal motion, for simplicity, at approximately twice the pendulum's natural frequency, that is,
$$\theta_d = 0.1 \sin(4\pi t) \ rad \tag{7}$$
which is at 2 Hz frequency (about twice the natural frequency of the uncontrolled system). The feedback control loop is performed at 1000 Hz sampling speed corresponding to a sampling period $T = 0.001s$. The desired frequency 2Hz is well below $\pi/2T$ and thus is a low frequency signal compared with sampling frequency. From the inset of figure 1, the magnitudes of ZOH and FOH at the desired frequency are 0.999993 and 1.000065 respectively, both of which are very close to 1. Thus the magnitude distortions are negligible. On the other hand, the phases of ZOH and FOH at the desired frequency are $-6\times10^{-3}$ and $-6\times10^{-7}$ respectively. The FOH has significantly less phase lag (about 10 to the 4th power less) than the ZOH does. Because of the excellent phase responses, the FOH discretization has been shown to increase the accuracy of the numerical model over more direct explicit methods in the real-time substructure testing [1].

A combination of feed-forward and feedback control is implemented on the pendulum. A DC servo-motor (Minertia Motor, FB5L20E) serves as the actuator while an optical encoder (with $0.09°$ resolution) is used to measure the pendulum angle, θ, from its equilibrium position [33, 34]. The control action is performed at 1000 Hz sampling rate on the pendulum that has a natural frequency of 1.1 Hz.

The linearized state space representation of the test setup is given as in [35, 36]

$$\dot{x} = Ax + \bar{B}V_a \tag{8}$$

$$x = \begin{bmatrix} \theta \\ \dot{\theta} \end{bmatrix}, \quad A = \begin{bmatrix} 0 & 1 \\ -\dfrac{mgl}{2J} & -\dfrac{b + \dfrac{K_bK_i}{R_a}}{J} \end{bmatrix}, \quad \bar{B} = \begin{bmatrix} 0 \\ \dfrac{K_i}{R_aJ} \end{bmatrix}$$

where $V_a$ is the control voltage (motor armature voltage) and the other parameters are electro-mechanical properties of the motor-pendulum assembly as listed in Table 1.

**Table 1.** Parameters of the experimental setup

| Parameter | Value | Unit |
|---|---|---|
| m (pendulum mass) | 0.125 | kg |
| l (pendulum length) | 0.33 | m |
| g (gravitational constant) | 9.807 | m/s² |
| J (rotational inertia) | 0.0042 | kgm² |
| $R_a$ (armature resistance) | 3.4 | Ω |
| $K_b$ (back-emf constant) | 0.0592 | Vs/rad |
| $K_i$ (torque constant) | 0.0592 | Nm/A |
| b (torsional damping coefficient) | 0.0045 | Nms/rad |

In order for the pendulum to follow the desired trajectory, a control structure shown in Figure 5 is implemented with the highlighted FOH sensing block

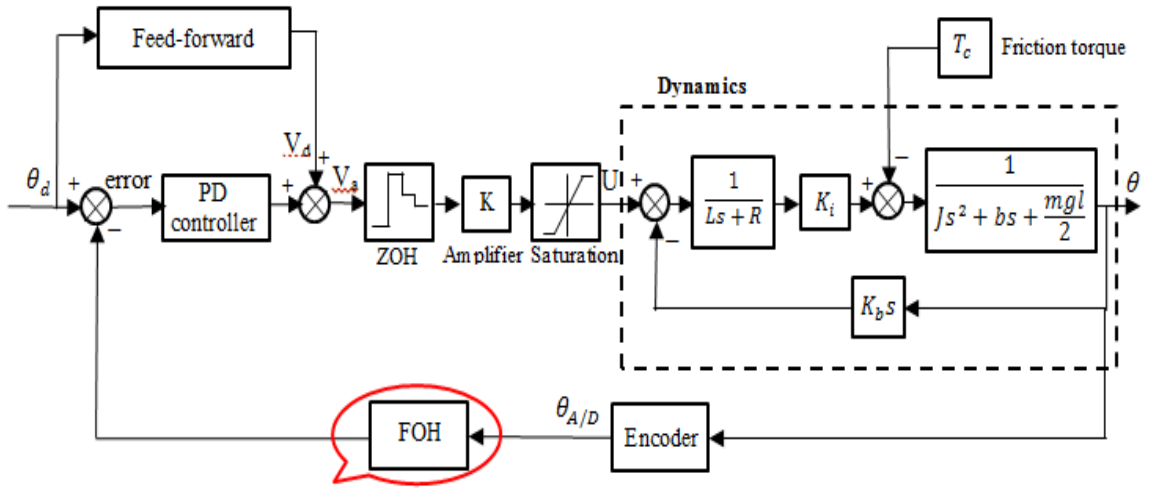

**Figure 5.** Block Diagram of the System Using the FOH

The feed-forward logic in the control is calculated as follows:

$$\dot{x}_d = Ax_d + \bar{B}V_d \tag{9}$$

$$V_d = \frac{R_a J}{K_i}\left(\ddot{\theta}_d + \frac{b + \frac{K_b K_i}{R_a}}{J}\dot{\theta}_d + \frac{mgl}{2J}\theta_d\right)$$

where $x_d = \begin{bmatrix}\theta_d \\ \dot{\theta}_d\end{bmatrix}$ is the desired trajectory and $V_d$ is the feed-forward control voltage. An important point to mention is that the amplitude of $\theta_d$ should be kept small in order to maintain the linearity in (8).

Subtracting (8) from (9) gives the error dynamics as

$$\dot{e} = Ae + \bar{B}(V_d - V_a) = Ae + \bar{B}u \qquad (10)$$

where $e = \begin{bmatrix}\theta_d - \theta \\ \dot{\theta}_d - \dot{\theta}\end{bmatrix}$ is the state vector describing the error, and $u = V_d - V_a = -Ke$ is the full state feedback control law.

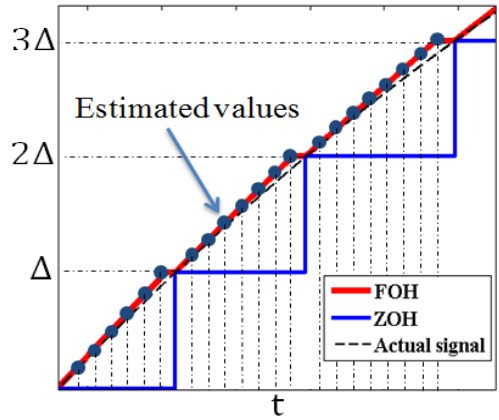

**Figure 6.** ZOH and FOH output signals

Figure 6 illustrates one actual signal with ZOH sampled signal and delayed FOH signal and it can be seen that the latter one yields a better approximation of the actual signal. As mentioned previously, the optical encoder with 4000 pulses per revolution has a sensor resolution $\Delta$ of 0.09 deg [37-40]. To estimate the analog signal between two quantized values, a first-order-hold (FOH) equivalent is applied to the ZOH signal $\theta_{A/D}$. The extrapolated signal is a piecewise linear approximation to the original analog signal that was sampled as shown in Figure 6. The slope of the previous step of the ZOH signal $\theta_{A/D}$ is used to estimate the output of the current step and the estimated value is obtained at the beginning of each sampling period. Since the FOH output is still not smooth enough (but yields much smaller errors in amplitude which is shown later), a second order low pass filter could be added to the FOH output.

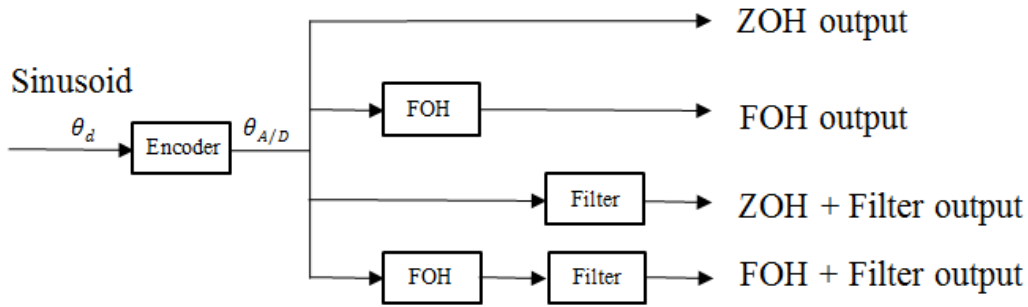
**Figure 7.** Comparison of various outputs with sinusoidal inputs.

Figure 7 shows various ways of sensing, i.e., ZOH, FOH, ZOH with filter and FOH with filter. In reality, a low-pass filter is usually used to eliminate high-frequency noise. The position of the FOH in the whole system is shown in Figure 5 and it is shown that the FOH is implemented for the signal obtained from the encoder. In order to compare the performance of the FOH output and ZOH output with/without the filter, a simulation is made to analyze the performance of the above methods on the sensing side as shown in Figure 7. The peak to peak errors between the various outputs and the sinusoidal input signal are obtained as listed in Table 2.

**Table 2.** Peak to peak error between the output of different sensing schemes and the sinusoidal input

| Outputs | Peak to peak error |
| --- | --- |
| ZOH | 1.9757% |
| FOH | 1.6986% |
| ZOH+Filter | 1.0135% |
| FOH+Filter | 0.8662% |

From Table 2, the FOH output yields smaller error than the ZOH output does. Also, adding a filter to the output yields apparent smaller errors than the corresponding original output. Out of all the listed methods, the filtered FOH output produces the best approximation to the continuous sinusoidal input. Based on this analysis, the closed-loop peak to peak errors with respect to different outputs on the sensing side are obtained on the simulation model (Figure 5). The highlighted FOH block is modified according to Figure 7 to get various outputs. The performance of different outputs from the peak to peak tracking error perspective is shown in the Table 3.

**Table 3.** Simulation result for the closed-loop peak to peak tracking error

| Outputs | Closed-loop peak to peak error |
|---|---|
| ZOH | 2.0286% |
| FOH | 1.5693% |
| ZOH+Filter | 1.0755% |
| FOH+Filter | 0.9303% |

The agreement between Table 3 and Table 2 shows that better sensing and reconstruction scheme yield smaller peak to peak tracking error. Finally, experimental results were done to verify the finding and show that the filtered FOH equivalence produces the best approximation to the continuous system out of all the methods examined (Table 4).

**Table 4.** Experimental result for the closed-loop peak to peak tracking error

| Outputs | Closed-loop peak to peak error |
|---|---|
| ZOH | 3.3404% |
| FOH | 2.7345% |
| ZOH+Filter | 2.2293% |
| FOH+Filter | 2.1790% |

Finally, the degree of the reduction of the closed-loop error for a simple trajectory tracking example is visualized in the discrete Fourier transformation (DFT) of the steady-state error, as depicted in Figure 8.

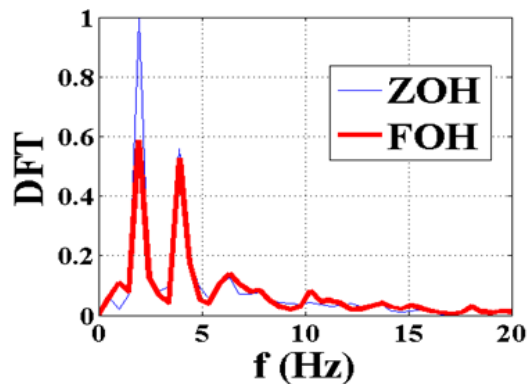

**Figure 8.** DFT of the closed-loop error using ZOH and FOH.

The scale of the vertical axis is normalized with respect to the maximum magnitude of the closed-loop error using ZOH sensing scheme, i.e. the ratio of $|Error_{ZOH}(\omega i)|/max|Error_{ZOH}(\omega i)|$ expressed in percent is shown in the figure. The light line represents the DFT of the steady-state error using the ZOH sensing scheme. The bold line depicts the DFT of the steady-state error using the FOH sensing scheme. It is observed that the dominant frequency component of 2 Hz (which is the desired frequency) is suppressed by about 40%, while the rest of the frequency spectrum remains practically unchanged.